\documentclass[fleqn,twoside,twocolumn,nofootinbib,showkeys]{revtex4} %
\usepackage[sec,nocpr,nopacs]{ujp_UTF8} 

\begin{document}
\title[Spectral Analysis and Invariant Measure]
{SPECTRAL ANALYSIS AND INVARIANT\\ MEASURE IN STUDIES OF THE
DYNAMICS\\ OF THE
HEMOSTASIS OF A BLOOD VESSEL}%
\author{V.~Grytsay}
\affiliation{\bitp}
\address{\bitpaddr}
\email{vgrytsay@bitp.kiev.ua}

\udk{577.3}  \razd{\seciii}

\autorcol{V.I.\hspace*{0.7mm}Grytsay}

\setcounter{page}{1}%

\begin{abstract}
A mathematical model of atherosclerosis of a blood vessel is
advanced with regard for the entry of low-density lipoproteins
(LDLs) into blood. For the first time, the influence of cytokines on
the inflammation of a blood vessel at the formation of
atherosclerotic plaques is taken into account. With the help of the
expansion in a Fourier series and the calculation of an invariant
measure, the scenario of the appearance of strange attractors
depending on a change in the parameter of dissipation of cholesterol
is studied. The conclusion is made about the interconnection of the
dynamics of the metabolic process in a blood vascular system and its
physical state.
\end{abstract}
\keywords{hemostasis, self-organization, strange attractor, Fourier
series, invariant measure, LDLs, cytokines.} \maketitle

\section{Introduction}

The blood composition stability has a particular meaning for the
vital activity of organism. There exist two counteracting systems in
blood vessels: namely, the fibrillation and liquefaction ones which
are in the functional equilibrium, by preventing the fibrillation of
blood and forming the hemostasis of a blood vessel \cite{1,2,3,4}.
The regulators of the fibrillation of blood are thromboxanes and
prostacyclins. A number of researchers modeled the given process and
other analogous ones, trying to get the comprehensive knowledge
about them \cite{5,6,7,8,9,10,11,12,13,14,15}. The author developed
a mathematical model of the prostacyclin-thromboxane system of a
blood vessel, determined the conditions for the self-organization of
the system and for the appearance of stationary stable autoperiodic
oscillations, and studied the conditions for the development of
hemophilia and thrombosis \cite{16,17,18,19,20,21}. In the frame of
this approach, a mathematical model of blood vessel with regard for
the presence of low-density lipoproteins (LDLs) such as ``poor
cholesterol'' in blood and the formation of atherosclerotic plaques
in blood vessels was constructed in \cite{22, 23}, and the influence
of LDLs on the self-organization of the hemostasis of a blood vessel
and the formation of dynamical chaotic modes in the system was
studied. The interval of a level of dissipation of cholesterol from
blood, where chaotic autooscillations can arise, was determined. The
phase-parametric diagram of autooscillatory modes depending on the
dissipation of cholesterol from blood was constructed. The scenario
of bifurcations with a period doubling until the aperiodic modes of
strange attractors arise due to the intermittence was given, and the
strange attractors appeared as a result of the formation of a mixing
funnel were constructed. The complete spectra of Lyapunov exponents
for various modes were calculated modes. For strange attractors, the
KS-entropy, ``foresight horizons,'' and Lyapunov's dimensions of the
fractality of attractors were determined. It was shown that the
cause for changes in the metabolic process of hemostasis of a blood
vessel can be the insufficient level of the dissipation of
cholesterol from blood. The study of the dependence of
autooscillatory modes on the fat concentration in blood was carried
out, and the chaotic modes of strange attractors were found. Such
modes lead to the disbalance between the amount of ``poor
cholesterol'' deposited in a blood vessel and its removal from the
system. This provokes the formation of plaques in arteries. It was
shown that LDLs affect the binding of thrombocytes and are deposited
in the walls of blood vessels. This causes the autocatalysis of
cholesterol in blood and an increase in its level. The mathematical
study of the determined modes was carried out.

In the present work, we continue the study of this system with regard for
the presence of ``poor cholesterol'' in blood. In more details, we will
consider the physical process of autocatalysis at the formation of
atherosclerotic plaques. For the first time, the mechanism of the given
process is modeled. According to author's hypothesis, it will be proved here
that the autocatalysis occurs due to the action of cytokines, and their role
in the given process will be estimated. Within the new model, we will study
the influence of the dissipation of cholesterol on the self-organization of
the metabolic process and on the appearance of a dynamical chaos in the
system. With the help of spectral analysis, by constructing the projections
of a phase portrait and its invariant measure, we will study the formation
of a strange attractor.

\section{Mathematical Model}

\begin{figure}[b]%
\vskip-3mm
\includegraphics[width=\column]{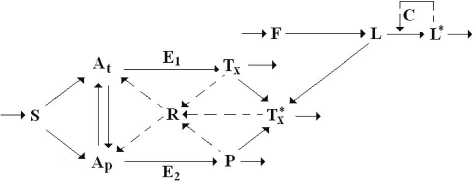}
\vskip-3mm\caption{General kinetic scheme of hemostasis of a blood
vessel }\vspace*{1.5mm}
\end{figure}

Mathematical model of atherosclerosis of the vessel: constructed by
the author according to the general scheme of hemostasis of the
blood vessel, taking into account the presence of "bad cholesterol"
in the blood [22,23]. In addition, this paper clarifies and expands
the overall scheme of the metabolic process of hemostasis of a blood
vessel. Instead of conditional enzymatic regulation of the process
of thrombosis ([22,23]), the metabolic process of cytokine influence
on the formation of atherosclerotic plaques is presented. . The
mathematical model of other metabolic processes remained unchanged,
with the corresponding parameters [22,23]. The general kinetic
scheme of hemostasis of a blood vessel is presented in Fig.1.

The mathematical model includes a system of 12 differential
equations (\ref{eq1})--(\ref{eq12}) describing changes in the
concentrations of components by the scheme shown in Fig.~1. In its
construction, we used the law of mass action and the enzyme
catalysis kinetics. The equations are written with regard for the
balance of masses of intermediate products of the reactions between
separate stages of the metabolic process. We have
\[
\frac{dA_t }{dt}=\frac{k_5 S}{(1+S+R^2)(1+k_6 T_x )}\,-
\]\vspace*{-7mm}
\begin{equation}
\label{eq1} -\, \frac{k_7 A_t E_1 }{(1+A_t +k_1 T_x )(1+E_1 )}+k_p
A_p -k_t A_t -\alpha _1 A_t ,
\end{equation}\vspace*{-7mm}
\begin{equation}
\label{eq2} \frac{dT_x }{dt}=\frac{k_7 A_t E_1 }{(1+A_t +k_1 T_x
)(1+E_1 )}-\frac{k_8 T_x^4 }{(k_9 +T_x^4 )}-\alpha _2 T_x ,
\end{equation}\vspace*{-10mm}
\[
\frac{dA_p }{dt}=\frac{k_2 SR^2}{(1+S+k_3 A_p )(k_4 +R^2)}\,-
\]\vspace*{-7mm}
\begin{equation}
\label{eq3} -\, \frac{k_{10} A_p E_2 }{(1+A_p )(1+E_2 )}+k_t A_t
-k_p A_p -\alpha _3 A_p ,
\end{equation}\vspace*{-7mm}
\begin{equation}
\label{eq4} \frac{dP}{dt}=\frac{k_{10} A_p E_2 }{(1+A_p )(1+E_2
)}-\frac{k_{11} T_x^\ast P^4}{(1+T_x^\ast )(k_{12} +P^4)}-\alpha _4
P,
\end{equation}\vspace*{-10mm}
\[
\frac{dE_1 }{dt}=\frac{k_{13} A_T }{(1+A_T )(1+R^4)}\, -
\]\vspace*{-7mm}
\begin{equation}
\label{eq5} -\, \frac{k_7 A_t E_1 }{(1+A_t +k_1 T_x )(1+E_1
)}-\alpha _5 E_1 ,
\end{equation}\vspace*{-7mm}
\[
\frac{dE_2 }{dt}=\frac{k_{15} A_p T_x^{\ast 4} }{(k_{16} +A_p
)(k_{17} + T_x^{\ast 4} )}\, -
\]\vspace*{-7mm}
\begin{equation}
\label{eq6} -\, \frac{k_{10} A_p E_2 }{(1+A_p )(1+E_2 )}-\alpha _6
E_2 ,
\end{equation}\vspace*{-7mm}
\begin{equation}
\label{eq7} \frac{dR}{dt}=k_{18} \frac{k_{19} +T_x^{\ast 4} }{k_{20}
+(T_x^\ast +k_{21} R)^4}-\alpha _7 R,
\end{equation}\vspace*{-7mm}
\begin{equation}
\label{eq8} \frac{dT_x^\ast }{dt}=k_8 \frac{L+T_x^4 }{k_9 +L+T_x^4
}-\frac{k_{11} T_x^\ast P^4}{(1+T_x^\ast )(k_{12} +P^4)}-\alpha _8
T_x^\ast .
\end{equation}\vspace*{-10mm}
\begin{equation}
\label{eq9} \frac{dF}{dt}=F_0 -l\frac{C}{1+C}\, \frac{F}{1+F+L},
\end{equation}
\begin{equation}
\label{eq10} \frac{dL}{dt}=k\frac{C}{1+C}\, \frac{F}{1+F+L}-\mu
\frac{LL^\ast }{1+L+L^\ast },
\end{equation}\vspace*{-5mm}
\begin{equation}
\label{eq11} \frac{dL^\ast }{dt}=\mu _1 \frac{LL^\ast }{1+L+L^\ast
}-\mu _0 L^\ast ,
\end{equation}\vspace*{-5mm}
\begin{equation}
\label{eq12} \frac{dC}{dt}=C_0 L^\ast \frac{F}{1+F}\,
\frac{N}{N+L}-\alpha _9 \, C.
\end{equation}
We consider that the principal regulators of the hemostasis and the
formation of thrombi are endotheliocytes of a blood vessel.

By Fig.~1, the input substance for the thrombosis-antithrombosis
system is essential arachidonic fat acid $S$ which enters blood from
intestinal tract. Under the action of phospholipases, the acid is
accumulated in thrombocytes $A_p $ (\ref{eq3}) and endotheliocytes
$A_t $ (\ref{eq1}) of a blood vessel. Then it is transformed by
prostaglandin-$H$-synthetase of thrombocytes $E_1 $ (\ref{eq5}) and
prostaglandin-$H$-synthetase of prostacyclins $E_2 $ (\ref{eq6})
with the formation of, respectively, thromboxanes $T_x $ (\ref{eq2})
and prostacyclins $P$ (\ref{eq4}). These prostaglandins are
antagonists to each other. By interacting, they form the dynamical
biochemical equilibrium in the fibrillation of blood. But if the
conditions of aggregation of thrombocytes are satisfied, then the
sequence of metabolic processes with their interaction starts. The
given reactions transform the input inflammatory substances that are
formed in atherosclerotic plaques at the aggregation of
thrombocytes~-- $T_x^\ast $ (\ref{eq8}). When the plaques are
destroyed, the inflammatory substances interact with blood and form
a thrombus, which leads to infarct or insult.

Depending on the formed levels of $P$, $T_x $, and $T_x^\ast $
(Fig.~1), the amount of the regulating component, cytoplasmatic
guanylate cyclase $R$ (\ref{eq7}), varies. It inhibits the
aggregation of thrombocytes. This occurs under the action of the
negative feedback affecting the level of activity of phospholipases
of endotheliocytes $A_t $ and thrombocytes $A_p $. Moreover, these
reactions regulate the entrance of arachidonic acid into
endotheliocytes and thrombocytes and the amounts of prostacyclin
(\ref{eq3}) and thromboxane (\ref{eq1}). A change in $R$ causes the
formation of different modes affecting the development of the
hemostasis of a blood vessel.

Another input substance of the system is fat molecules $F$
(\ref{eq9}). They are transported by blood along arteries and
influence the level of ``poor cholesterol,'' i.e., LDL. Its level is
denoted by the variable $L$ (\ref{eq10}). It is formed in liver and
small intestine.

From blood, the excess of LDL particles enters the arterial wall of
a blood vessel, chemically changes, and accumulated there. Modified
LDLs stimulate endotheliocytes to the activation of adhesion
receptors which bind to monocytes of blood and lure them into
intima. In addition, monocytes produce numerous mediators of
inflammation, in particular, cytokines (transmitters of signals
between cells of the immune system) C (\ref{eq12}), and cover their
surface by waste receptors that help them to absorb modified LDLs.

In intima, monocytes ripen and become active mac\-ro\-pha\-ges.
Macrophages absorb LDLs, by filling themselves with fat and forming
a foam of oxidized lipoproteins. The fat strips which are the early
manifestation of atherosclerotic plaques $L^\ast $ (\ref{eq11}) are
formed.

Cytokines and inflammatory molecules formed in foamy cells of
macrophages favor the further growth of plaques.

In the metabolic process, the positive feedback of the growth of
plaques $L^\ast $ and the concentration of inflammatory substances
$T_x^\ast $ is formed, because cytokines C activate
(\ref{eq10})--(\ref{eq12}) the process of absorption of LDLs
(\ref{eq10}), (\ref{eq11}) by macrophages.

The accumulation of cholesterol in arteries and the growth of
plaques cause thrombophilia, the narrowing of the blood-bearing
channel of an artery, and stenosis.

Depending on the dissipation of LDLs in the me\-ta\-bo\-lic process
of hemostasis of a blood vessel, the autooscillatory and chaotic
modes can appear instead of stationary ones.

The model contains the following parameters:
\[
k=4; \quad k_1 =3; \quad k_2 =1; \quad k_3 =5; \quad k_4
=10;\]\vspace*{-6mm}
\[
k_5 =2.1; \quad k_6 =5; \quad k_7 =2; \quad k_8 =1.5; \quad k_9
=5;\]\vspace*{-6mm}
\[
 k_{10} =0.75; \quad k_{11} =0.3; \quad k_{12} =15; \quad k_{13} =0.75;\]\vspace*{-6mm}
\[
 k_{15} =1; \quad k_{16} =0.5; \quad k_{17} =5; \quad k_{18} =5;\]\vspace*{-6mm}
\[
 k_{19} =0.02; \quad k_{20} =25; \quad k_{21}
=0.5; \quad k_p =0.1;\]\vspace*{-6mm}
\[
k_t =0.1; \quad S=2; \quad \alpha _1 =0.01; \quad \alpha _2
=0.01;\]\vspace*{-6mm}
\[
\alpha _3 =0.01; \quad \alpha _4 =0.173; \quad \alpha _5 =0.05;
\quad \alpha _6 =0.07;\]\vspace*{-6mm}
\[
\alpha _7 =0.2; \quad \alpha _8 =0.0021; \quad \alpha _9 =0.2; \quad
F_0 =0.01;\]\vspace*{-6mm}
\[
l=2; \quad \mu =4; \quad \mu _0 =0.42; \quad \mu _1
=2.3;\]\vspace*{-6mm}
\[
C_0
=11; \quad N=0.05.
\]

\begin{figure*}%
\vskip1mm
\includegraphics[width=15.8cm]{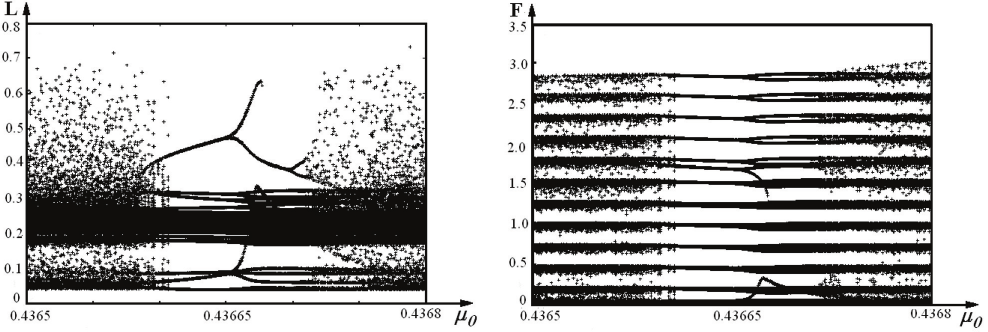}\\
{\it a\hspace*{8.2cm}b} \vskip-3mm\caption{Phase-parametric diagrams
of the variables $L(t)$ and $F(t)$ versus the parameter of the
dissipation of cholesterol from a blood vessel $\mu _0 $  }
\end{figure*}

\begin{figure*}%
\vskip1mm
\includegraphics[width=15.8cm]{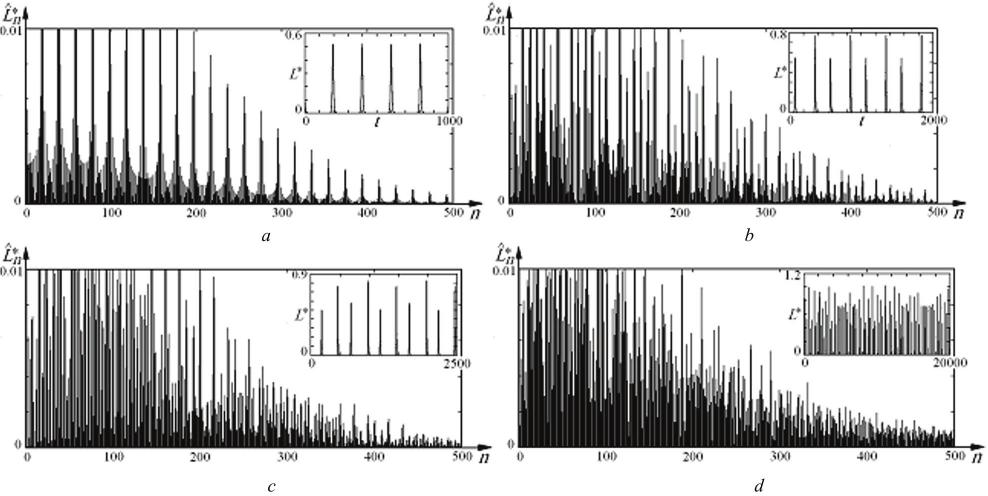} \vskip-3mm\caption{Distribution of Fourier harmonics in the modes of successive
period doubling and dynamical chaos of the metabolic process in a
blood vessel: regular attractor $2^0$ $(\mu _0 =0.43)$ [22]
(\textit{a}) ; regular attractor $1\times 2^1$ $(\mu _0 =0.435)$
[22] (\textit{b}); regular attractor $1\times 2^2$ $(\mu _0
=0.43555)$ [22] (\textit{c}); strange attractor $1\times 2^x$ $(\mu
_0 =0.437)$ [22] (\textit{d})}\vspace*{-2mm}
\end{figure*}

\begin{figure*}%
\vskip1mm
\includegraphics[width=13.5cm]{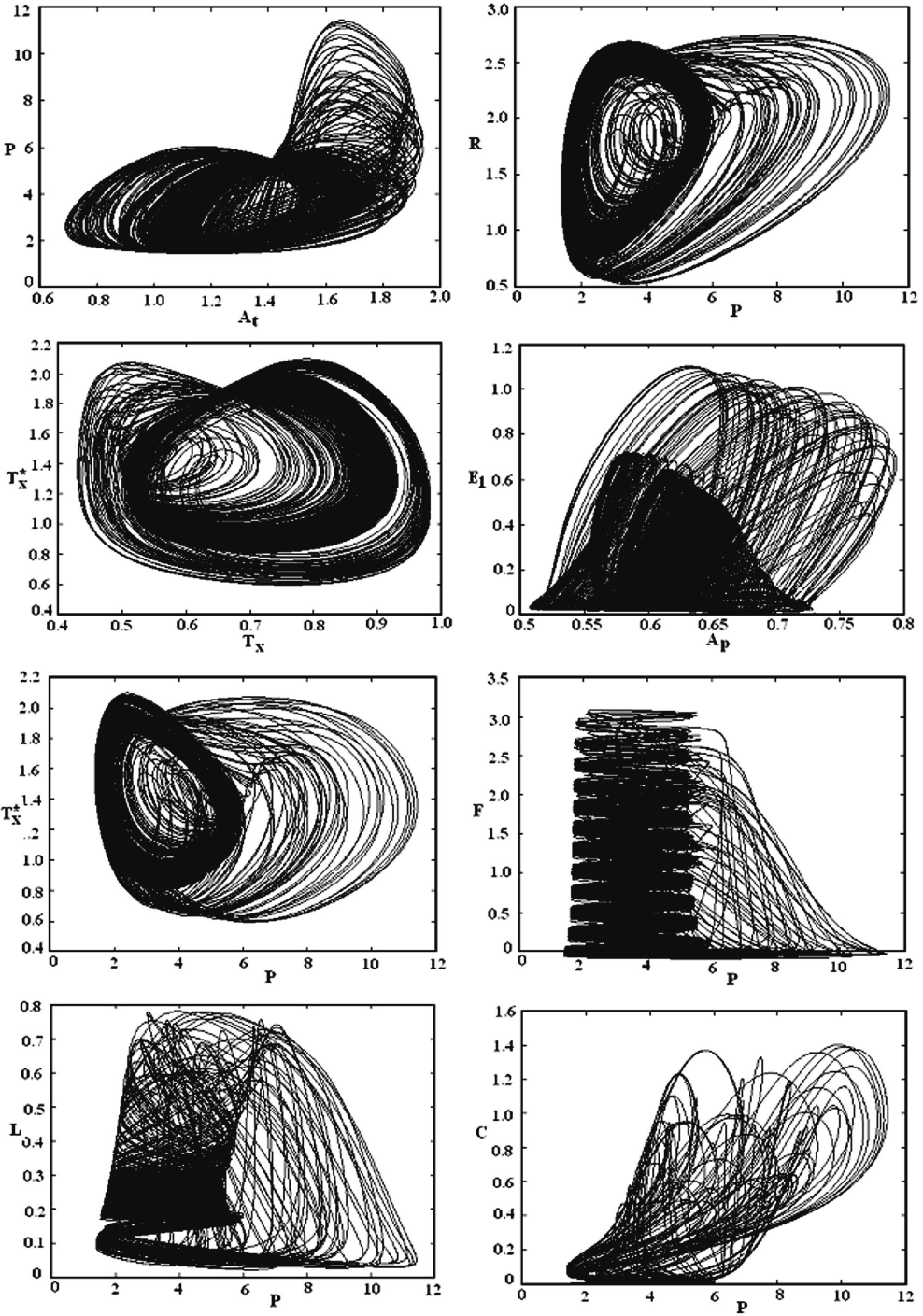} \vskip-3mm\caption{Projections of the phase portrait of a strange attractor
$1\times 2^x$ at $\mu _0 =0.437$:  in the plane ($A_t $, $P$)
(\textit{a}); in the plane ($P$, $R$) (\textit{b}); in the plane
($T_x $, $T_x^\ast $) (\textit{c}); in the plane ($A_P $, $E_1 $)
(\textit{d}); in the plane ($P$, $T_x^\ast $) (\textit{e}); in the
plane ($P$, $F$) (\textit{f}); in the plane ($P$, $L$) (\textit{g});
in the plane ($P$, $C$) (\textit{h})}
\end{figure*}

\begin{figure*}%
\vskip1mm
\includegraphics[width=13.5cm]{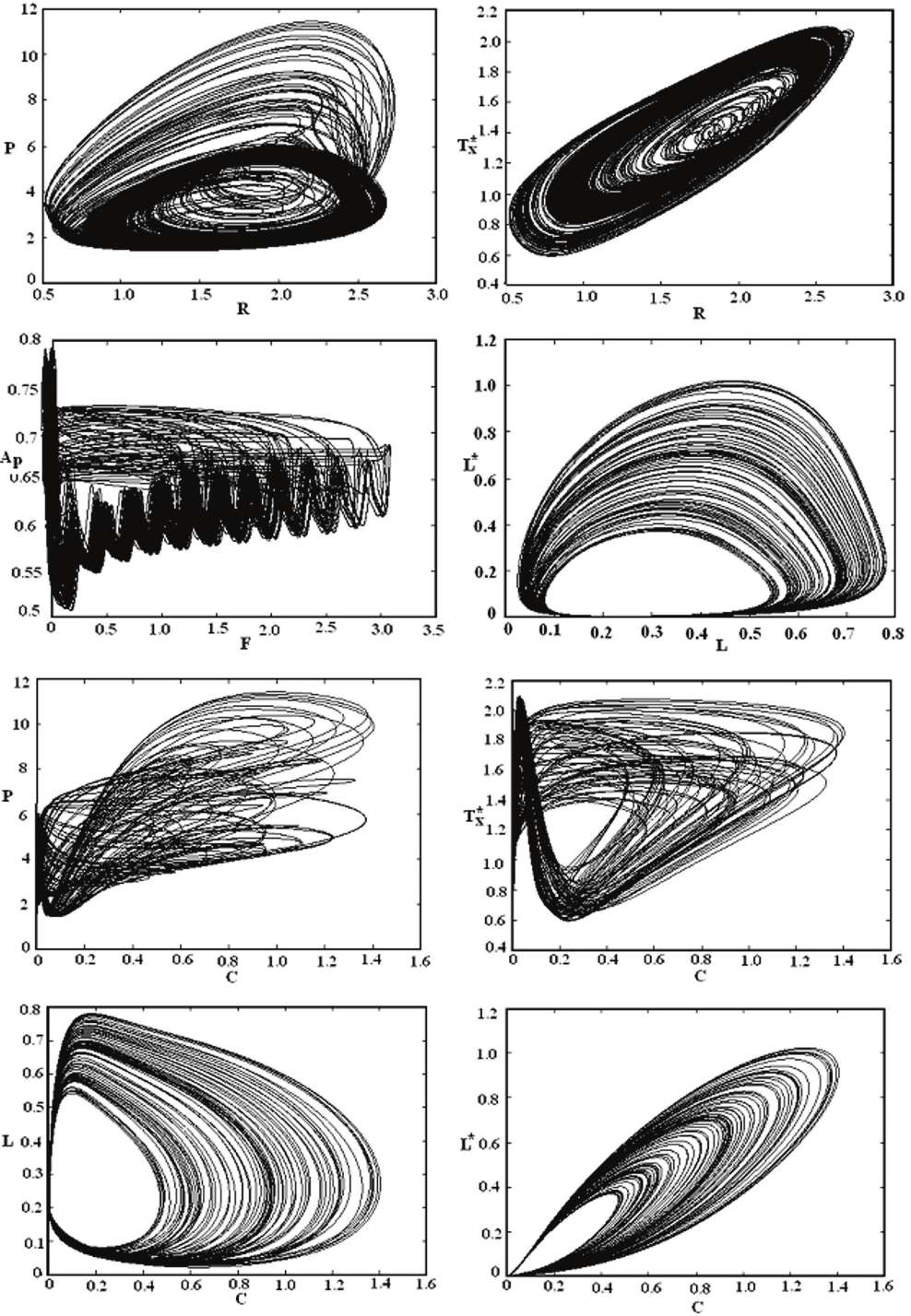} \vskip-3mm\caption{Projections of the phase portrait of a strange attractor
$1\times 2^x$ at $\mu _0 =0.437$: in the plane ($R$, $P$)
(\textit{a}); in the plane ($R$, $T_x^\ast $) (\textit{b}); in the
plane ($C$, $P$) (\textit{c}); in the plane ($L$, $L^\ast $)
(\textit{d}); in the plane ($P$, $T_x^\ast $) (\textit{e}); in the
plane ($C$, $T_x^\ast $) (\textit{f}); in the plane ($C$, $L$)
(\textit{g}); in the plane ($C$, $L^\ast $) (\textit{h})}
\end{figure*}
\begin{figure*}%
\vskip1mm
\includegraphics[width=14.3cm]{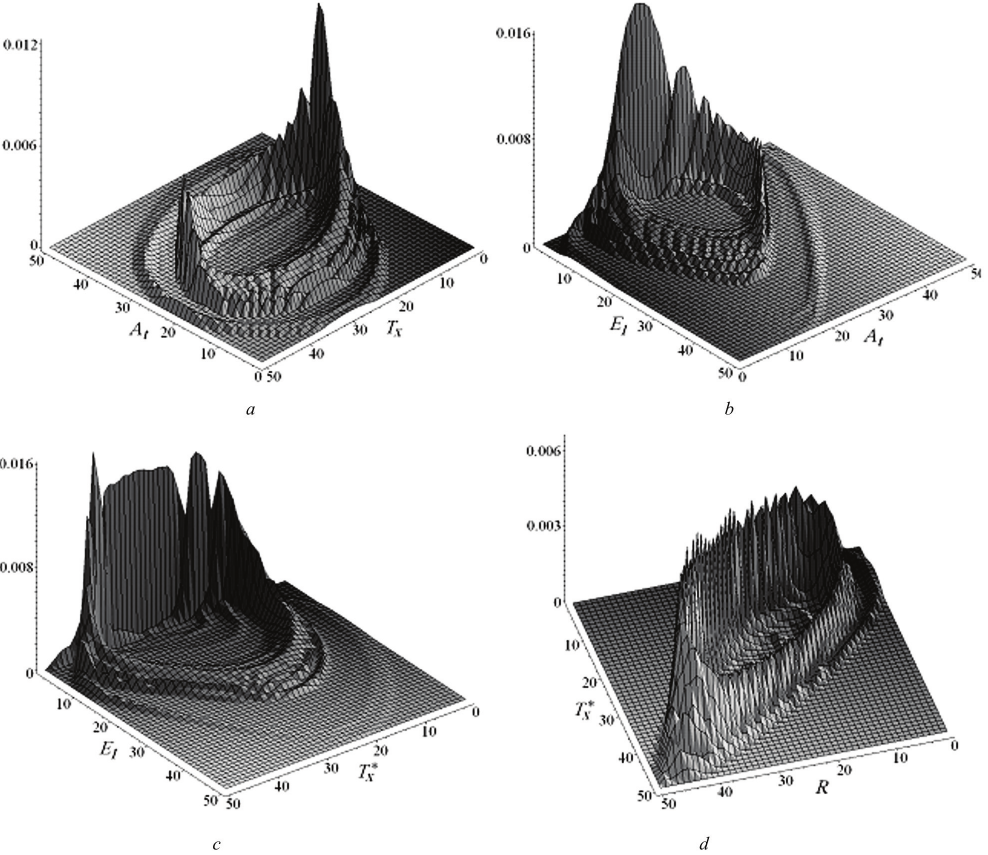} \vskip-3mm\caption{Histograms of projections of the invariant measure of the
strange attractor $1\times 2^x$, $\mu _0 =$0.437: on the plane $(A_t
,T_x )$ (\textit{a}); on the plane $(E_1 ,A_t )$ (\textit{b}); on
the plane $(E_1 ,T_x^\ast )$ (\textit{c}); on the plane $(R,T_x^\ast
)$ (\textit{d})}
\end{figure*}

\noindent The study of solutions of the mathematical model
(\ref{eq1})--(\ref{eq12}) was executed with the help of the
application of the theory of nonlinear differential equations [24,
25] and the developed methods of mathematical modeling of
biochemical systems [26--59]. The numerical solution was carried out
by the Runge--Kutta--Merson method with an accuracy of $10^{-8}$.

\section{Results of Studies}\vspace*{-1.5mm}

In the present work, we continue the investigation of the
autooscillatory modes in the metabolic process of hemostasis of a
blood vessel. The obtained dynamical modes [22,23] indicate the
dependence of the dynamics of the process on a change in the level
of cholesterol in blood.

We now consider the influence of cholesterol plaques on the
appearance of a dynamical chaos in the given process.

The calculations showed that, in the interval $\eta _0 \in
(0.43,  0.438)$, the transition from the stable autoperiodic modes
to chaotic ones occurs in the system.

In Fig.~2,~\textit{a},~\textit{b}, we present the phase-parametric
diagrams of the system for the variables $L(t)$ and $F(t)$ as
functions of the parameter $\mu _0 $ corresponding to the
dissipation of cholesterol from a blood vessel.

In the construction of the phase-parametric diagrams, we apply the
method of cutting [22].

We now study the scenario of the transition from the autooscillatory
modes to a dynamical chaos with the help of spectral analysis. In
Fig.~3, we give the spectra of the expansion of the kinetic curve
$L^\ast $ in a Fourier trigonometric series in the modes of
successive period doubling. The plots of this curve on the
corresponding time interval are shown in the upper right corners. In
the construction of the spectral patterns, we took 1000 harmonics.

As the coefficient $\mu _0 $ increases, the multiplicity of the
autoperiodic process grows as well. According to the doubling of the
multiplicity of the autoperiodic process, the number of basic peaks
of harmonics increases two times
(Fig.~3,~\textit{a},~\textit{b},~\textit{c}.), and, eventually, the
harmonics of turbulence appear (Fig.~3,~\textit{d}). The given
chaotic mode of the strange attractor $1\times 2^x$ arises by the
Feigenbaum scenario due to the intermittence at $\mu _0 =0.437$
[22].

In Figs. 4 and 5, we present various projections of the mode of a
strange attractor at $\mu _0 =0.437$. It is seen that the given
strange attractor is formed due to the funnel. There, we observe the
mixing of trajectories which approach one another along some
directions, but diverge along other ones. At a slight fluctuation
due to the intermittence, the periodic process becomes unstable, and
a determinate chaos arises.

Then we calculated the invariant measure of the obtained strange
attractor. According to the Krylov--Bogolyubov theorem, at least one
invariant measure exists in the case of a continuous mapping of the
compact phase space of a dynamical system. Based on the calculated
invariant measure of a strange attractor, we calculated and drew the
histograms of projections of this measure on the corresponding
planes (see Fig.~6). These histograms correspond to projections of
the phase portrait (Figs. 4 and 5) and demonstrate clearly the
course of a trajectory of the dynamical system in the phase space.

\section{Conclusions}

With the help of an improved mathematical model, further study of
the effect of ``bad cholesterol'' on the metabolic process of
hemostasis of a blood vessel.

For the first time the mechanism of influence of cytokines on
formation of atherosclerotic plaques of a vessel is modeled. Using
the model, it is theoretically proven that this leads to
autocatalysis of the growth of atherosclerotic plaques and the
progressive development of atherosclerosis.

The metabolic process of thrombosis and antithrombosis systems is
considered as a dissipative system, the stationary dissipative
structure of which is hemostasis of a blood vessel. The presence of
``bad cholesterol'' and cytokines unbalances this process. They are
autocatalysts of blood levels and the formation of self-oscillating
regimes in hemostasis. Spectral analysis revealed a scenario of
bifurcations doubling the period of self-oscillations, until
eventually due to discontinuity in the formation of the mixing
funnel chaotic regimes of strange attractors. Various projections of
the phase portrait of the strange attractor and histograms of
projections of its invariant measure are constructed. The obtained
results allow to investigate the influence of low-density
lipoproteins in the self-organization of the metabolic process of
hemostasis of a blood vessel. It has been shown that cytokines
formed in atherosclerotic plaques can cause thrombosis at any time,
anywhere in a blood vessel.

\vskip3mm \textit{The present work was partially supported by the
Program of Fundamental Research of the Department of Physics and
Astronomy of the National Academy of Sciences of Ukraine
``Mathematical models of non-equilibrium processes in open systems''
No.\,0120U100857. }

\rezume{%
В.\,Грицай}{СПЕКТРАЛЬНИЙ АНАЛІЗ\\ ТА ІНВАРІАНТНА МІРА ПРИ ВИВЧЕННІ\\
ДИНАМІКИ ГЕМОСТАЗУ КРОВОНОСНОЇ СУДИНИ} {В даній роботі продовжується
дослідження математичної моделі атеросклерозу кровоносної судини  з
урахуванням  надходження  в  кров ліпопротеїдів низької щільності
(ЛПНЩ). В даній моделі вперше враховано вплив цитокінів на запалення
судини при утворенні атеросклеротичних бляшок.  За допомогою
розкладу в ряд Фур'є та розрахунку інваріантної міри досліджено
сценарій виникнення дивних атракторів в залежності від зміни
параметра дисипації холестерина.  Зроблено висновки про
взаємозв'язок  динаміки метаболічного процесу кровоносної системи і
її фізичним станом.}{\textit{Ключові слова:} гемостаз,
самоорганізація, дивний атрактор, ряд Фур'є, інваріантна міра, ЛПНЩ,
цитокіни.}

\end{document}